# Multi-View Depth Consistent Image Generation Using Generative AI Models

*Application on architectural design of university buildings*

Xusheng Du[1], Ruihan Gui[2], Zhengyang Wang[1], Ye Zhang[2], and Haoran Xie[1]

[1]*Japan Advanced Institute of Science and Technology.*
[2]*Tianjin University.*
[1]*s2320034@jaist.ac.jp, 0009-0008-1086-2081*

**Abstract.** In the early stages of architectural design, shoebox models are typically used as a simplified representation of building structures but require extensive operations to transform them into detailed designs. Generative artificial intelligence (AI) provides a promising solution to automate this transformation, but ensuring multi-view consistency remains a significant challenge. To solve this issue, we propose a novel three-stage consistent image generation framework using generative AI models to generate architectural designs from shoebox model representations. The proposed method enhances state-of-the-art image generation diffusion models to generate multi-view consistent architectural images. We employ ControlNet as the backbone and optimize it to accommodate multi-view inputs of architectural shoebox models captured from predefined perspectives. To ensure stylistic and structural consistency across multi-view images, we propose an image space loss module that incorporates style loss, structural loss and angle alignment loss. We then use depth estimation method to extract depth maps from the generated multi-view images. Finally, we use the paired data of the architectural images and depth maps as inputs to improve the multi-view consistency via the depth-aware 3D attention module. Experimental results demonstrate that the proposed framework can generate multi-view architectural images with consistent style and structural coherence from shoebox model inputs.

**Keywords.** Multi-view consistency, image generation, generative AI models, architectural design, university buildings

## 1. Introduction

The rapid progress of generative AI, particularly deep learning-based generative models, has shown great potential in various research fields and applications including architectural design (Li et al., 2024). Generative AI models are designed to be capable of automatically generating photorealistic architectural images with high quality renderings, significantly reducing the labour-intensive process of transforming design concepts into detailed visual representations. These models provide powerful support for architects to efficiently propose and validate design ideas and are becoming essential emerging solutions in the field of architecture design (Almaz et al., 2024).

In the early stages of architectural design, shoebox models effectively communicate initial design concepts by abstracting building structures into simplified 3D models, providing clearer spatial and structural information than sketches or text. However, transforming shoebox models into detailed designs is often complex and time-consuming. As shown in Figure 1, our method aims to simplify the workflow by using generative AI models to directly generate design renderings from shoebox models.

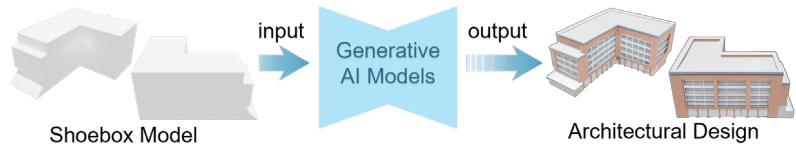

*Figure 1. Research Purpose*

For generating architectural images from shoebox models, it is a crucial task to conduct consistent image generation for multi-view inputs. We aim to generate images of the same building object from different perspectives while ensuring image consistency in geometric structure and style. This consistency between different views may affect both the visual quality of the design and the accurate representation of spatial layouts and structural relationships. This issue is particularly important for multi-story buildings with complex structures to ensure design accuracy and uniformity. However, current generative AI models often fail to maintain multi-view consistency, leading to the deviation from the initial design concepts.

To address these issues, we propose a novel three-stage multi-view depth-consistent image generation approach. The proposed method employs generative models in conjunction with a multi-view consistency mechanism, with a particular focus on image generation for multi-story university buildings. We choose university buildings as generation targets because they present unique challenges, such as maintaining the alignment of floors and preserving stylistic features such as the position of windows and details on the facade. Our method enables the generation of multi-view designs with consistent style, structure, and depth from simple shoebox model inputs.

The main contributions of this work are summarized as follows:

- We propose a three-stage multi-view image generation framework that enables the generation of architectural designs from simple shoebox model inputs.
- We introduce an image space loss module that complements the latent space loss to improve the structural accuracy and stylistic consistency of multi-view images.
- We construct a publicly accessible dataset of university teaching buildings that includes shoebox model images paired with corresponding architectural designs.

## 2. Related Work

### 2.1. MULTI-VIEW CONSISTENCY

# MULTI-VIEW DEPTH CONSISTENT IMAGE GENERATION USING GENERATIVE AI MODELS

Achieving multi-view consistency is a core challenge in AI-driven multi-view image generation and 3D reconstruction. While there are currently limited studies specifically addressing architectural image consistency, there has been significant research on models and algorithms for multi-view consistency. These studies lay the foundation for addressing this challenge in architectural design. Zero123++ (Shi et al., 2023) employs conditional generative diffusion models to produce multi-view consistent images from single-view inputs, effectively mitigating inconsistencies across perspectives. MVControl (Li et al., 2024) achieves precise viewpoint control through its hybrid diffusion prior, supporting geometric and structural consistency in multi-view geometry and structural consistency in multi-view outputs. DreamComposer (Yang et al., 2024) can generate consistent images by leveraging multi-view features extracted from 3D representations, making it suitable for multi-view tasks that require controlled viewpoint and style consistency. While existing methods effectively address multi-view consistency challenges in general contexts, they have not yet been extensively applied to architectural image generation. This study aims to build on these advances to address specific architectural challenges such as structural accuracy, style coherence, and multi-view consistency in complex building designs.

## 2.2. MULTI-MODAL FUSION

The use of multi-modal information such as RGB images and depth maps can significantly improve multi-view consistency, especially when generating complex 3D scenes or architectural structures. Depth maps provide essential geometric constraints that help to improve multi-view consistency. MiDaS (Dueholm et al., 2022), a monocular depth estimation tool, strengthens structural consistency in multi-view generation tasks by generating multi-view depth maps. MVD-Fusion (Hu et al., 2024) employs a multi-view depth fusion framework that combines RGB images and depth maps to control geometric consistency in 3D reconstruction tasks. One-2-3-45 (Liu et al., 2024) combines depth information with viewpoint conditioning to produce multi-view consistent 3D models from single-view images. SyncDreamer (Liu et al., 2023) leverages a 3D-aware feature attention mechanism to synchronize the intermediate states of multi-view images, further ensuring consistency across generated views. Inspired by them, this study aims to further integrate depth and RGB data to improve multi-view consistency in architectural image generation.

## 3. Method

We propose a three-stage framework that progressively improves the structural and style consistency of the generated multi-view architectural images as shown in Figure 2. Firstly, an improved multi-view diffusion model is proposed to generate intermediate architectural images from shoebox model inputs. Depth information is then estimated from the intermediate images using monocular depth estimation method. Finally, the depth information is integrated with the intermediate architectural images through a multi-view fusion process.

## 3.1. MULTI-VIEW DIFFUSION MODEL

The proposed generation framework takes shoebox model images from the early architectural design stage as input, and aims to generate multi-view architectural designs with color, texture, and structural details as output. We improve ControlNet (Zhang et al., 2023) model by using a multi-view module to achieve multi-view diffusion generation. Unlike single view to multiple image generation, this module can process multiple viewpoints simultaneously and ensure consistency across views.

As shown in the first stage in Figure 2, the multi-view ControlNet model is improved based on the stable diffusion model (Rombach et al., 2022), which optimizes the single-view diffusion process into a multi-view generation module. To improve the multi-view consistency, we incorporate a cross-attention mechanism. By embedding text features into noisy latent features and sharing features from different views, the generated design can more accurately follow the textual prompts (such as style description, architectural requirements, etc.), improving the generation quality and maintaining structural consistency and visual continuity across different views.

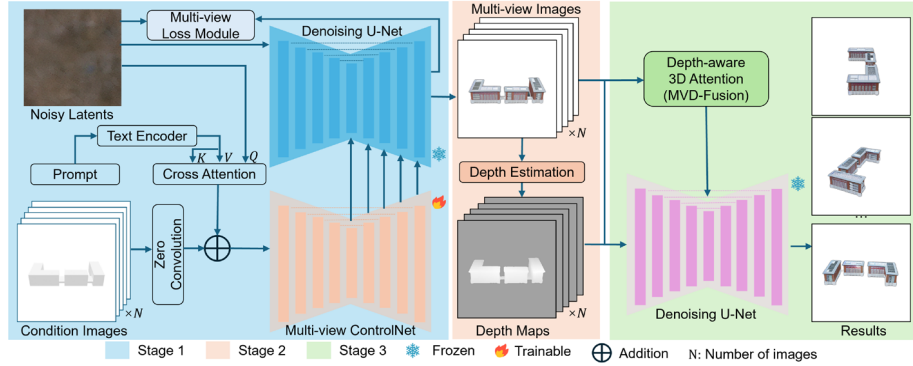

*Figure 2. Overview of the Three-Stage Framework*

To optimize multi-view consistency in the generated images, in addition to the basic loss in the ControlNet latent space, we design and incorporate a loss module in the image space as shown in Figure 3. This module further improves multi-view consistency through three additional cross-view constraint losses: style consistency loss, structural consistency loss, and angle alignment consistency loss.

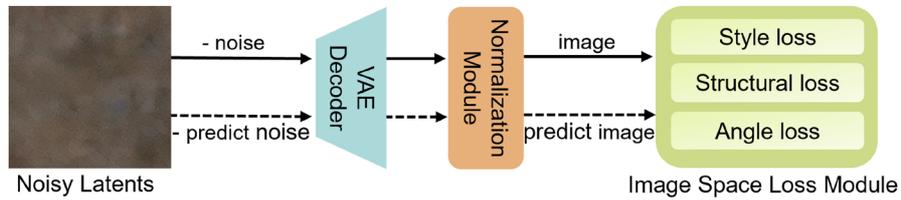

*Figure 3. The Proposed Multi-View Consistency Losses for Generative Model*

The consistency of the style between the generated and the reference image is evaluated by calculating the difference in the Gram matrices of each viewpoint based on the Visual Geometry Group (VGG) (Simonyan and Zisserman, 2015) feature map. This loss ensures that the images generated from all viewpoints exhibit the same style.

# MULTI-VIEW DEPTH CONSISTENT IMAGE GENERATION USING GENERATIVE AI MODELS

$$L_{style} = \sum_{i=1}^{N} ||G(\phi(g_i)) - G(\phi(r_i))||^2$$

Where $N$ is the number of viewpoints. $g_i$ and $r_i$ are the generated and real images in the same viewpoint. $G(\cdot)$ is the Gram matrix for capturing style features. $\phi(g_i)$ and $\phi(r_i)$ are the feature maps extracted by VGG network.

The consistency of the geometric structure is guaranteed by two distinct types of loss: perceptual loss and multi-view content consistency loss. Perceptual loss compares the VGG features of the generated image and the reference image.

$$L_{percep} = \sum_{i=1}^{N} ||\phi(g_i) - \phi(r_i)||^2$$

The multi-view content consistency loss compares the feature differences between the generated image and the reference image in neighbouring viewpoints to ensure the content consistency of the generated image in multiple viewpoints.

$$L_{content\_cos} = \sum_{i=1}^{N-1} (||\phi(g_i) - \phi(g_{i+1})||^2 - ||\phi(r_i) - \phi(r_{i+1})||^2)^2$$

Angle alignment consistency loss compares the pixel differences between the generated image and the real image in adjacent viewpoints to ensure the consistency of the generated image and the real image in terms of geometric information across multiple viewpoints.

$$L_{angle\_cos} = \sum_{i=1}^{N-1} (||g_i - g_{i+1}||^2 - ||r_i - r_{i+1}||^2)^2$$

$$L_{total} = \alpha L_{style} + \beta L_{percep} + \gamma L_{content\_cos} + \delta L_{angle\_cos}$$

Where $\alpha, \beta, \gamma, \delta$ denote the loss weights. The total loss optimizes style, structure, and angle consistency across views during training, resulting in architectural designs that maintain consistent appearance and structural features across multiple views.

## 3.2. DEPTH ESTIMATION

Depth information estimation in real-world datasets often exhibit inconsistencies due to differences in data sources and acquisition methods. In addition, the inherent diversity of architectural designs, including variations in geometric structure and style, further complicates the challenge of maintaining consistency in depth information. To address the challenges posed by inconsistencies and biases in available depth data for architectural design, our framework adopts a depth estimation strategy based on multi-objective learning. This approach enables the integration of depth information from multiple datasets while mitigating the depth inconsistency problem. By improving structural consistency across different views, the depth estimation strategy supports multi-view coherence for diverse architectural designs.

In our implementation, we utilize MiDaS (Dueholm et al., 2022), a state-of-the-art monocular depth estimation tool, as a pre-trained model for extracting depth information. Depth features are extracted from the architectural designs generated in the first stage by employing a encoder with multi-scale feature extraction capabilities. These features are captured from each view of the design to generate high quality depth maps as complementary information to improve the structural consistency of the generated results.

### 3.3. MULTI-VIEW FUSION

In the third stage, the generated images and corresponding depth maps of each view are fed into MVD-Fusion (Hu et al., 2024), a multi-view depth consistency fusion framework, to verify and optimize the design with consistent style and structure. By employing 3D reconstruction method based on multi-view depth consistency, this fusion mechanism effectively reduces depth differences between different views, and minimizes deformation issues during generation, and achieves robust feature fusion across multiple fields of view. Combining this multi-view fusion mechanism with feature alignment ensures the structural and stylistic consistency of the reconstructed 3D model, we can achieve the coherent multi-view presentation of the building.

Note that the main feature of the proposed method is adopting additional depth-aware cross-attention layers in the U-Net framework of ControlNet (Zhang et al., 2023). These attention layers are distributed throughout the various layers of the U-Net model, capturing cross-view depth features during encoding and decoding to optimize perceptual feature consistency among multi-view images. The residual cross-attention layer dynamically maps features from different views into a shared depth latent space. The cross-attention mechanism can process multi-view input in parallel, learning the feature relationships between different views to represent the geometric information of the building structure consistently across views.

## 4. Experiments

### 4.1. TRAINING DATASET

We constructed a specialised dataset to generate detailed architectural designs from architectural shoebox models. We began by conducting an extensive survey of university teaching buildings in China, analyzing their architectural characteristics and features. Based on this survey, we collected models of university buildings and captured their essential structural elements to create simplified shoebox models for these buildings. This effort resulted in a comprehensive paired 3D dataset containing 210 shoebox models and their corresponding detailed models.

Given the large amount of data that needs to be processed when working with 3D models, we decoupled the multi-view design generation task into a simple image-to-image generation task. Based on the initial dataset, we used Blender to render from multiple angles and generate a multi-view image dataset for training and testing. To include the roof features, a 30-degree perspective angle was used, with a 6-degree rotation applied to generate 60 views per model. A total of 12,600 images of the shoebox models and 12,600 images of the building model details were obtained.

### 4.2. IMPLEMENTATION DETAILS

The model was trained on a high-performance computing server equipped with an NVIDIA A100 GPU (80GB) and CUDA 12.5, to meet the computational demands of large-scale multi-view architectural generation tasks. The loss weights α, β, γ, and δ were set to $10^9$, 100, 1, and 10, respectively, to balance the importance of each loss term. The training took approximately 200 hours and around 500,000 iterations.

MULTI-VIEW DEPTH CONSISTENT IMAGE
GENERATION USING GENERATIVE AI MODELS

### 4.3. RECONSTRUCTION & GENERATION EXPERIMENTS

After completing the model training, we conducted two types of experiments, reconstruction experiments and generation experiments, to comprehensively evaluate the model's performance. The reconstruction experiments aim to verify the adaptability of the model to the training data. We selected a set of architectural shoebox model images from the training dataset and fed them into the model to generate corresponding architectural design images. The reconstruction results are shown in Figure 4. The generation experiments focus on evaluating the model's ability to generalize to unseen data. We used architectural shoebox model images that never appeared in the training process as input and used the model to generate corresponding architectural design images. The generation results are shown in Figure 5.

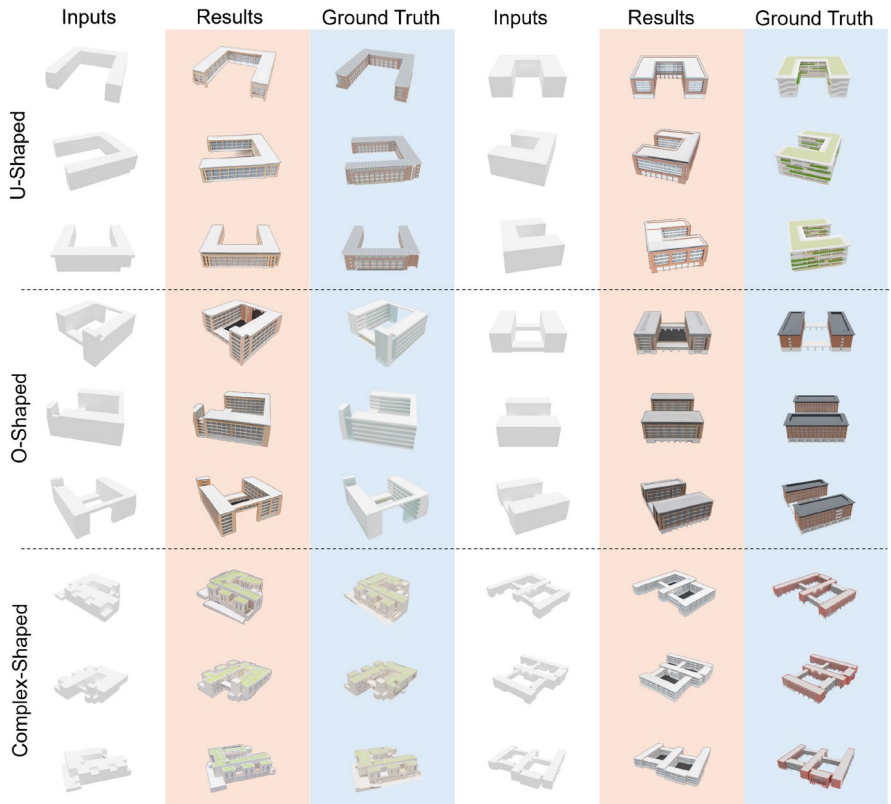

*Figure 4. Reconstruction Results for U-, O- and Complex-Shaped University Buildings*

### 4.4. USER EVALUATION

To evaluate the performance and practical applicability of our model in generating consistent and high-quality architectural designs, we conduct the user evaluation with 15 graduate students (9 male and 6 female) from the Department of Architecture. Each participant was presented with 14 reconstructed models and the corresponding 14 target models, as well as 10 generated models, each with 6 images representing different perspectives. The study assessed the effectiveness of the model-generated architectural designs in real-world applications across six core dimensions: structural integrity, structural consistency, detail integrity, detail consistency, visual aesthetics, and practicality. Participants used a 5-point scale (1 for strongly disagree, 5 for strongly agree) to rate the generated images on these dimensions. The analysis of the user evaluation results is shown in Figure 6.

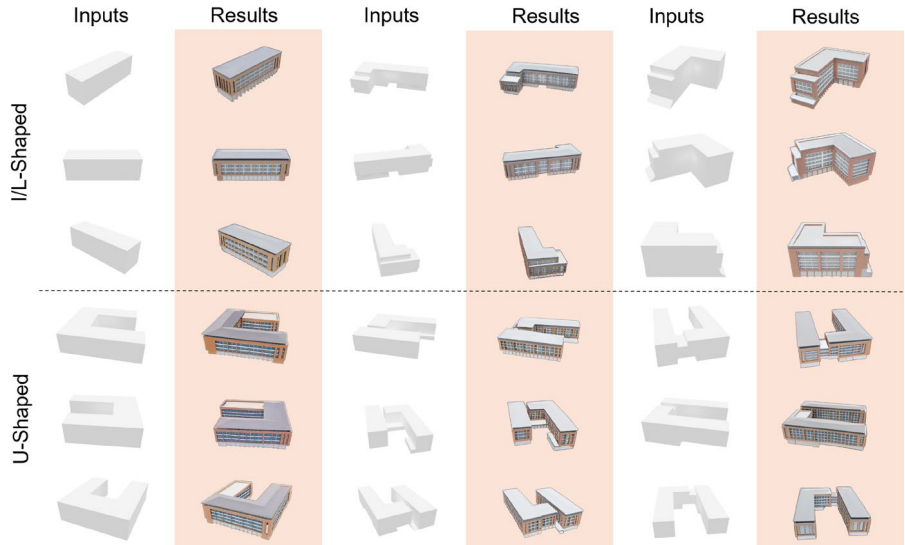

*Figure 5. Generation Results for I/L-Shaped and U-Shaped University Buildings*

## 5. Results

### 5.1. RECONSTRUCTION RESULTS

Figure 4 shows the reconstruction results, we found that the proposed model can accurately recover the detailed features of the architectural design. The structural geometric consistency of the reconstruction is well preserved. In addition, the model excels at capturing textural details, providing a realistic and coherent visual output. These results highlight the effectiveness of the model in dealing with complex architectural structures such as multi-storey facades and the joining of multiple shoebox models, while maintaining accuracy from different perspectives. This reconstruction capability highlights the adaptability of the model to training data and its ability to produce high quality output for architectural visualisation tasks.

### 5.2. GENERATION RESULTS

Figure 5 shows the results of the generation experiment, which shows that the model can generate realistic and consistent images of architectural designs, and has strong generalisation capabilities for unknown data. The multi-view images generated are consistent in style and structure, and excel in the geometric alignment of architectural structures and the rendering of texture details.

# MULTI-VIEW DEPTH CONSISTENT IMAGE GENERATION USING GENERATIVE AI MODELS

## 5.3. USER EVALUATION RESULTS

Figure 6 shows the statistical analysis of the evaluation results. For the reconstruction results, the average scores for structural integrity (M=3.567, SD=1.234) and structural consistency (M=3.533, SD=1.324) indicate a solid level of multi-view stability and accuracy. Detail integrity (M=3.557, SD=1.099) and detail consistency (M=3.390, SD=1.033) also show reliable and consistent performance. The generation results, while slightly lower, still show promising results. Structural integrity (M=3.553, SD=1.257) and structural consistency (M=3.433, SD=1.303) show comparable stability and accuracy to the reconstruction results, highlighting the robustness of the model even with unseen inputs. For detail integrity (M=3.353, SD=1.126) and detail consistency (M=3.280, SD=1.190), the results suggest that the model maintains a reasonable level of detail accuracy, although there is room for further refinement in fine-grained features. In terms of visual aesthetics and practicality, both reconstruction and generation results scored slightly lower than the structure and detail metrics, but still indicate moderate success. Reconstruction results achieved mean scores of M=3.167 (SD=1.035) for visual aesthetics and M=3.424 (SD=1.128) for practicality, reflecting a good balance between visual appeal and functional usability. Generation results, with mean scores of M=3.027 (SD=1.083) for visual aesthetics and M=3.373 (SD=1.099) for usability, highlight the potential for improvement while demonstrating the model's ability to produce usable outputs even for novel inputs.

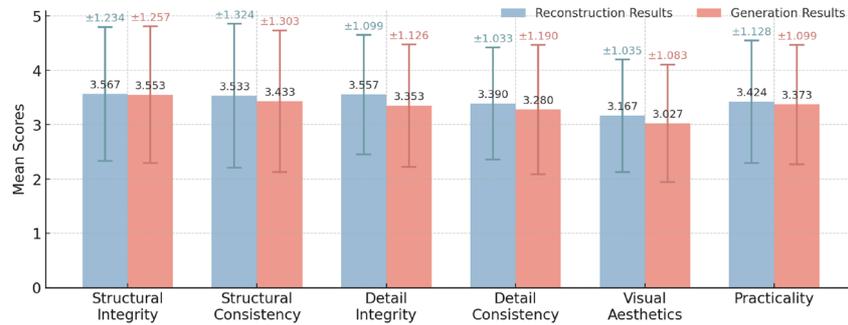

*Figure 6. User Evaluation Results*

## 6. Conclusion

This work proposed a novel multi-view consistent image generation method based on generative AI models, focusing on the generation of multi-view consistent architectural renderings from the common shoebox model representation. Experimental results show that the proposed method has significant advantages in terms of multi-view consistency, detailed integrity and overall generation quality. However, the proposed method also has some limitations. First, the shoebox model used in the experiment is too simple a case and cannot fully reflect the variety and complexity of real-world architectural input. When dealing with more complex and realistic input data, this limitation may hinder the model's ability to generalize effectively. As shown in Figure 7 (a), the extracted shoebox model does not represent the hollow structure in the middle, and the shoebox model in Figure 7 (b) does not represent the structure between the stairs well, resulting in a mismatch between the generated results and the target structure. Future work will focus on integrating more diverse input datasets beyond university models, exploring a broader array of architectural styles, facade patterns, and materials, and improving the model's ability to control fine-grained details. Through these extensions, we aim to transition beyond the proof-of-concept stage and develop solutions that can be effectively adopted in real-world architectural production settings.

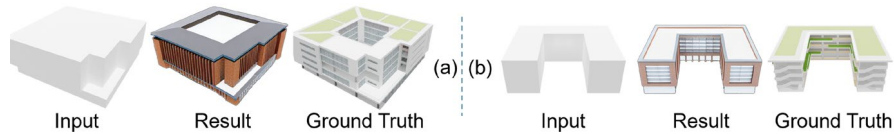

*Figure 7. Examples of Structural Mismatch in Reconstruction*